\documentclass[aps,prl,amsmath,amssymb,preprint,nofootinbib%
]{revtex4-1}
\usepackage[latin9]{inputenc}
\usepackage{amsmath}
\usepackage{amssymb}
\usepackage{graphicx}
\usepackage{verbatim}
\usepackage{bm}
\usepackage{color}

\newcommand{\nn}{\nonumber}

\newcommand{\sgn}{\operatorname{sgn}}

\newcommand{\la}{\langle}
\newcommand{\ra}{\rangle}
\newcommand{\rar}{\rightarrow}

\newcommand{\be}{\begin{eqnarray}}
\newcommand{\ee}{\end{eqnarray}}

\newcommand{\bs}{\begin{equation}\begin{split}}
\newcommand{\es}{\end{split}\end{equation}}

\begin{document}
\title{Spin Correlations in Quantum Wires}
\author{Chen Sun}
\affiliation{Department of Physics, Texas A\&M University, College Station, Texas 77843-4242, USA}
\author{Valery L. Pokrovsky}
\affiliation{Department of Physics, Texas A\&M University, College Station, Texas 77843-4242, USA}
\affiliation{Landau Institute for Theoretical Physics, Chernogolovka, Moscow District, 142432, Russia}

\date{\today}

\begin{abstract}
We consider theoretically spin correlations in an 1D quantum wire with Rashba-Dresselhaus spin-orbit interaction (RDI). The correlations of non-interacting electrons display electron-spin resonance at a frequency proportional to the RDI coupling. Interacting electrons on varying the direction of external magnetic field transit from the state of Luttinger liquid (LL) to the spin density wave (SDW) state. We show that the two-time total spin correlations of these states are significantly different. In the LL the projection of total spin to the direction of the RDI induced field is conserved and the corresponding correlator is equal to zero. The correlators of two components perpendicular to the RDI field display a sharp ESR driven by RDI induced intrinsic field. In contrast, in the SDW state the longitudinal projection of spin dominates, whereas the transverse components are suppressed. This prediction indicates a simple way for experimental diagnostic of the SDW in a quantum wire.
\end{abstract}
\maketitle

\textit{Introduction.---}Abanov et al. \cite{spin resonance} predicted a sharp ESR in a quantum wire driven by an intrinsic momentum dependent RDI of the spin-orbit-interaction origin. The electron interaction in \cite{spin resonance} was neglected. In the work \cite{spin resonance in LL} theory of the RDI spin resonance was extended to the interacting 1D electron system described as the Luttinger liquid (LL).\cite{Tomonaga,Luttinger,Luttinger correction} Theory \cite{spin resonance in LL} showed that the ESR persists in the electronic LL with slightly modified shape of line. However, Starykh et al. \cite{Starykh} predicted that a quantum wire with the RDI subject to external magnetic field  perpendicular to the intrinsic field develops an SDW with the wave vector $2k_F$, an inhomogeneous state with physical properties rather different from the LL.

Multiple experiments on 1D wires (c.f. the review \cite{experiment review} and its references) have proved validity of the LL model. However, neither the ESR nor the SDW state was observed experimentally in quantum wires with the RDI. The standard way of the resonance observation, i.e. resonance absorption of an electromagnetic wave, gives a very weak signal since the total number of electrons in a wire is typically small. Therefore, it requires enormous power of the incident wave in the terahertz range of frequencies at helium temperature.\cite{spin resonance} The standard method of the SDW observation in the bulk is neutron scattering which doesn't work well for wires for the same reason. Purely magnetic measurements also are impossible. Here we propose to overcome these difficulties by using the spin noise measurements method developed by Crooker et al.\cite{Crooker 2004} In this method they measured total spin fluctuations in real time observing induced by them Faraday rotation of the light polarization. They were able to see them in quantum dots \cite{Crooker 2010,Crooker} that also contain not too large number of electrons. Therefore, their method is specially intended for weak signals. It does not require e-m sources of ultra-high power. The recently developed ultrafast spin noise spectroscopy have achieved frequency resolutions up to hundreds of Ghz.\cite{Hubner}

In this work we calculate spin correlators in the 1D interacting electron system with the RDI for both the ordinary LL and the SDW states. Different properties of spin correlations for these two states make it possible to identify them experimentally.

\textit{Model.---}We consider a quantum wire with the RDI in single channel regime, in which the electrons occupy only the lowest band of the transverse motion. Thus, the system is effectively 1D. We also assume zero temperature and the thermodynamic limit, i.e. wire's length being much larger than all other length scales. In 1D the most general form of the RD Hamiltonian is $H_{RD}=\alpha (\hat{n}\cdot \boldsymbol{\sigma}) p$, where $\alpha$ is the RDI coupling constant with dimensionality of velocity, $\hat{n}$ is a unit vector in the spin space, $\boldsymbol \sigma$ is the vector of Pauli matrices, and $p$ is the 1D momentum. (Later we set $\hbar=1$ and write everywhere wave vector $k$ instead of momentum $p$.) We choose the spin-orbit axis to be $z-$axis and assume the external magnetic field to lie in the $xOz$ plane: $\boldsymbol{b}=\frac{1}{2}g\mu_B \boldsymbol{B}=b_\parallel \hat z+b_\perp \hat x$. The Hamiltonian reads:
\begin{align}\label{}
H=\sum_{i}[\frac{k_i^2}{2m}+(\alpha k_i-b_\parallel)\sigma_{i,z}-b_{\perp} \sigma_{i,x}]+H_{int},
\end{align}
where $i$ labels electrons and $H_{int}$ denotes the e-e interaction. We assume external field to be weak: $b \ll\alpha k_F$, and the RDI velocity being much smaller than the Fermi velocity: $\alpha \ll v_F$. The time-ordered spin correlators are defined as:
\begin{equation}\label{definition of ST}
S^T_{aa}(t)=\la \operatorname {T_t} S_a(t)S_a(0)\ra=\la e^{iH|t|}S_a e^{-iH|t|}S_a\ra,
\end{equation}
where $S_a(t)$ is the total spin projection operator along direction $a=x, y, z$, and $\la\ra$ denotes the average over the ground state. The retarded correlators can be obtained from the time-ordered ones as $S^R_{aa}(t)=-2\Theta(t)\operatorname{Im}S^T_{aa}(t)$ \cite{1d book}, where $\Theta(t)$ is the Heaviside step function.

Let us analyze the symmetry of our model. The e-e interaction is invariant under any spin rotation. Thus, in the absence of the RDI and external field the Hamiltonian is $SU(2)$ invariant and all three components of the total spin are conserved. A finite RDI coupling $\alpha$ and/or parallel field $b_\parallel$ reduces the symmetry group to $U(1)$, in which case only $S_z$ is conserved. In the presence of non-zero transverse field $b_\perp$, the $U(1)$ symmetry is also broken and neither of the total spin components is conserved.

\textit{Ideal 1D Fermi gas.---}Before calculating the spin correlators of interacting electrons, it is instructive to solve the same problem for the ideal 1D Fermi gas. In the presence of the RDI and the external magnetic field, the spectrum consists of a pair of asymmetric parabola with avoided crossing. If $b \ll\alpha k_F$ and $\alpha \ll v_F$, the four Fermi momenta are approximately $k_{\sigma\tau}=\tau k_F-\sigma m[\alpha-\tau\frac{b_\parallel}{k_F}+\frac{b_\perp^2}{2k_F(\alpha k_F-\tau b_\parallel)}]$, where $\sigma=\pm$ denotes the spin-up/down bands and $\tau=\pm$ denotes right/left movers. At $T=0$ spin correlators can be obtained directly by calculating the ground state average:
\begin{align}\label{3}
&S^R_{xx}(t)=S^R_{yy}(t)=\Theta(t)\frac{2l}{\pi\alpha t}\sin(2\alpha k_Ft)\sin(2m\alpha^2t),\quad S^R_{zz}(t)=0,
\end{align}
where $l$ is the wire's length, and spins are in units $\hbar/2\equiv1/2$. The Fourier transforms are:
\begin{align}\label{non-int}
&S^R_{xx}(\omega)=S^R_{yy}(\omega)=\frac{ l}{2\pi\alpha}\log\frac{(\omega+i\delta)^2-[2\alpha(k_F+m\alpha)]^2}{(\omega+i\delta)^2-[2\alpha(k_F-m\alpha)]^2}, \quad S^R_{zz}(\omega)=0,
\end{align}
where $\delta=0^+$. The $x$ and $y$ components are equal and the $z$ component vanishes, respecting the $U(1)$ symmetry. The imaginary part of $S^R_{xx}(\omega)$ has a narrow peak around $\omega = 2\alpha k_F$ of the width $4m\alpha^2$. In the resonance interval of frequency $2\alpha(k_F-m\alpha)<|\omega|<2\alpha(k_F+m\alpha)$, the absorption intensity $\Im S_{xx}^{R}$ is constant. Disorder can change this exotic shape of line.

\textit{Interacting electrons and bosonization.---}For interacting electrons we apply Luttinger liquid theory, detailed description of which can be found in \cite{1d book, bosonization book}. In 1D the interaction between fermions near Fermi points is always strong enough to destroy the Fermi-excitations. Instead the Bose-excitations play the role of almost free quasiparticles. Before translation to the bosonic language (bosonization), the original quadratic spectrum of fermions is linearized around the Fermi points, and the infinite sea of negative energy levels is filled. The extension of the fermion spectrum to $-\infty$ contradicts to the initial spectrum of fermions limited from below. Usually this does not lead to mistakes in physical results if the substantial range of momenta is close to the Fermi points. However in some problems a broader range of momentum is important. Namely this happens in the case of the total spin correlators as it will be shown later. In this situation the LL theory can be used only together with a proper cut-off of negative momenta.

Below we consider the bosonization for the fermions with the RDI in a parallel external field. In terms of the fermionic particle field $\Psi_{\sigma}(x)$, $H_{int}$ reads: $H_{int}=\frac{1}{2}\sum_{\sigma,\sigma'}\int dx dx'U(x-x')\Psi^\dag_{\sigma}(x)\Psi^\dag_{\sigma'}(x')\Psi_{\sigma'}(x')\Psi_{\sigma}(x)$. We do not specify a form of the e-e interaction, except that it is assumed to be repulsive and short-ranged. The field $\Psi_{\sigma}(x)$ is the sum of the right and left movers fields $R_\sigma$ and $L_\sigma$. These chiral fields are expressed in terms of bosonic fields in a way that respects fermionic anticommutation relations. Details of the bosonization procedure are presented in the Supplemental Material \cite{sm}. The resultant Hamiltonian in terms of the bosonic charge fields $\phi_{c}$, $\theta_{c}$ and spin fields $\phi_{s}$, $\theta_{s}$ reads:
\begin{align}\label{18}
&H=H_c+H_s,\nn\\
&H_c=\frac{1}{2}\int dx[\frac{v_c}{K_c}(\partial_x\phi_c)^2+v_c K_c(\partial_x\theta_c)^2],\nn\\
&H_s=\frac{1}{2}\int dx[\frac{v_s}{K_s}(\partial_x\phi_s)^2+v_s K_s(\partial_x\theta_s)^2]+H_C,
\end{align}
where $K_c=\{1+[2\tilde U(0)-\tilde U(2k_F)]/(\pi v_F)\}^{-\frac{1}{2}}$, $v_c=v_F/K_c$, $K_s=[1-\tilde U(2k_F)/(\pi v_F)]^{-\frac{1}{2}}$, $v_s=v_F/K_s$ are the Luttinger parameters, and $H_C=\frac{g_C}{2(\pi a_0)^2}\int dx \cos(\sqrt{8\pi}\phi_s-\frac{4 b_\parallel}{ v_F} x)$ with $g_C=\tilde U(2k_F)$. ($\tilde U(q)$ is the Fourier transform of $U(x)$.) We approximated $\tilde{U}(2(k_F+\frac{\sigma b_\parallel}{ v_F}))$ by $\tilde U(2(k_F))$, and ignored a term mixing charge and spin fields $\propto \frac{b_\parallel}{\epsilon_F}\ll 1$. The charge and spin degrees of freedom in this Hamiltonian are separated. The charge Hamiltonian $H_c$ is completely quadratic, but the spin Hamiltonian $H_s$ contains a cosine term $H_C$. If $H_C$ can be neglected, the remaining quadratic Hamiltonian $H_s$ describes the ordinary LL state. When $H_C$ dominates, $\phi_s$ field becomes pinned to one of the minima of cosine, resulting in ordering in the spin sector -- the SDW state.

\textit{SDW in weak magnetic field.---}Starykh et al. \cite{Starykh} proved that static SDW appears when external field is directed perpendicular to the internal one and strongly exceeds it. We consider a more realistic limit $b\ll\alpha k_F$, and first fix $b_\perp=0$. The charge Hamiltonian $H_c$ is quadratic and does not change in magnetic field. To renormalize the spin part we define following \cite{Starykh} spin currents: $\vec{J}_R=\sum_{\nu,\nu'=\mp}R_\nu^\dag\frac{\vec{\sigma}_{\nu\nu'}}{2}R_{\nu'}$, $\vec{J}_L=\sum_{\nu,\nu'=\mp}L_\nu^\dag\frac{\vec{\sigma}_{\nu\nu'}}{2}L_{\nu'}$.
In terms of these currents, $H_s$ reads:
\begin{align}
&H_s=2\pi v_s' \int dx\{(J_R^zJ_R^z+J_L^zJ_L^z)+y_sJ_R^zJ_L^z\nn\\
&+y_C[(\cos\frac{4b_\parallel}{ v_F}x)(J_R^xJ_L^x+J_R^yJ_L^y)+(\sin\frac{4b_\parallel}{ v_F}x)(J_R^xJ_L^y-J_R^yJ_L^x)]\},
\end{align}
where $v_s'=\sqrt{2v_s^2(0)-v_F^2}=v_F[1-\tilde{U}(2k_F)/(2\pi v_F)]$, and the initial values of coupling constants are $y_s(0)=y_C(0)=-\tilde{U}(2k_F)/(\pi v_s')$. The constants $K_s$, $g_C$ and $y_s$, $y_C$ are connected by relations: $K_s=\sqrt{(2-y_s)/(2+y_s)}$, $g_C=-\pi v_s'y_C$. At $b_\parallel=0$, $H_s$ reduces to:
$H_s=2\pi v_s' \int dx[(J_R^zJ_R^z+J_L^zJ_L^z)+y_sJ_R^zJ_L^z+y_C(J_R^xJ_L^x+J_R^yJ_L^y)]$. The renornalization group (RG) equations for the vertices $y_s$ and $y_C$ in one-loop approximation read: $dy_s/d\lambda=y^2_C$, $dy_C/d\lambda=y_sy_C$, where the running RG parameter is $\lambda=\log(l_r/a_0)$ and $l_r$ is the running scale of length. The integral of motion $y_C^2-y_s^2=y_C^2(0)-y_s^2(0)=0$ implies that the RG flow goes along the separatrix to the fix point $y_s=y_C=0$. Thus, at large scales $K_s\rar 1$ and $g_C\rar 0$. The renormalization of $g_C$ to zero means irrelevance of $H_C$. The renormalization of $K_s$ to 1 demonstrates the $SU(2)$ invariance, since the RDI can be removed by a unitary transformation\cite{spin resonance in LL}. Therefore, at zero field the Hamiltonian is renormalized to a completely quadratic one. No SDW appears, and the system remains in an ordinary LL state with $K_s=1$. In finite $b_\parallel$ %the parallel external field
the SDW state also does not exist. The parallel field violates the $SU(2)$ symmetry leaving only the $U(1)$ symmetry.
In this case the term $H_C$ develops an oscillating factor $\exp(i\frac{4b_\parallel}{ v_F}x)$. For a more general case including also $b_\perp$, it is modified to $\exp[i\frac{4b_\parallel}{ v_F}(1-\frac{b_\perp^2}{2\alpha^2 k_F^2-b_\parallel^2})x]$. The component $b_\perp$ enters only as a higher order correction. Due to oscillation the renormalization stops at a scale $l_0=\frac{ v_F}{4b_{\parallel}}$ determined by the strength of $b_\parallel$ %the parallel field
rather than the size of the system ($l$). For the parallel external field $\sim100 $Gs, $l_0$ is of the order of several micrometers. (In numerical estimates we use the data for In$_{0.53}$Ga$_{0.47}$As; see \cite{spin resonance} for references). Thus, in the thermodynamic limit $l\gg l_0$, any weak parallel field destroys the SDW. The coupling constants are not renormalized, and $H_C$ can be neglected. In completely perpendicular field the SDW should exist as proved in \cite{Starykh}. Thus, in weak field limit, SDW appears only if the external field is \textit{completely} perpendicular. The possible states of the wire are summarized in Table 1. But they may be different at stronger %perpendicular
field $b\gtrsim\alpha k_F$. Starykh et al. \cite{Starykh} considered opposite limiting case $b\gg \alpha k_F$ and argued that a weak parallel field does not destroy SDW.

\bigskip{}
\makebox[0.9\textwidth]{Table 1. States of the wire at $b\ll\alpha k_F$.}

\smallskip{}
\makebox[0.9\textwidth]{%

\begin{tabular}{c|c|c}
\hline
external field  & renormalization  & state\tabularnewline
\hline
$b=0$  &  $y_s\rar 0,y_C\rar 0$  & ordinary LL \tabularnewline
$b_\parallel\ne 0$  & none  & ordinary LL\tabularnewline
$b_\parallel=0,b_\perp\ne0$  & $y_s\rar -\infty,y_C\rar -\infty$ &  SDW\tabularnewline
\hline
\end{tabular}%
} \bigskip{}

\textit{Spin density correlations.---}Below we calculate the spin density correlators for the ordinary LL state and the SDW state. At $b\ll \alpha k_F$ the Fermi momenta are approximately $k_{\sigma\tau}=\tau k_F-\sigma m\alpha$. In the ordinary LL state, the cosine term $H_C$ can be dropped and the Hamiltonian becomes completely quadratic. The Luttinger parameters are given in the text following Eq. \eqref{18}, except $K_s=1$ at zero external field. Spin density operators read: $s_a(x)=\Psi^{\dag}_{\sigma}(x)\sigma_{a,\sigma\sigma'}\Psi_{\sigma'}(x)$, where $a=x,y,z$. The time-ordered spin density correlators are $s_{aa}(x,t)=\la \operatorname{T_t} s_a(x,t)s_a(0,0)\ra$. Applying the bosonization one can express spin correlators as path integrals over bosonic fields. Details of calculation are placed in \cite{sm}. The results are:
\begin{align}\label{dc for oLL}
&s_{xx}(x,\tau)=s_{yy}(x,\tau)=\frac{a_0^{K_s+\frac{1}{K_s}-2}}{\pi^2}\frac{(y_s^2-x^2)\cos(2m\alpha x )}{(x^2+y_s^2)^{1+\frac{K_s}{2}+\frac{1}{2K_s}}}+\frac{a_0^{K_c+\frac{1}{K_s}-2}}{\pi^2}\frac{\cos(2k_Fx)\cos(2m\alpha x )}{(x^2+y_c^2)^{\frac{K_c}{2}}(x^2+y_s^2)^{\frac{1}{2K_s}}},\nn\\
&s_{zz}(x,\tau)=\frac{K_s}{\pi^2}\frac{y_s^2-x^2}{(x^2+y_s^2)^2}+\frac{a_0^{K_c+K_s-2}}{\pi^2}\frac{\cos(2k_Fx)}{(x^2+y_c^2)^{\frac{K_c}{2}}(x^2+y_s^2)^{\frac{K_s}{2}}},
\end{align}
where $y_{s/c}(\tau)=v_{s/c}\tau$, $\tau$ is imaginary time, and $a_0$ is an ultraviolet cut-off. Each correlator contains contributions from small $q$ and from $q\sim2k_F$. For weakly interacting case $K_c, K_s\approx1$, and both decay as $x^{-2}$ and oscillate.

The SDW state exists at completely perpendicular field, when $y_C$ flows to the strong coupling limit $y_C\rar -\infty$. $H_C$ is relevant and dominates the spin Hamiltonian. The field $\phi_s$ is pinned to $\phi_s=(N+\frac{1}{2})\sqrt{\frac{\pi}{2}}$ ($N$ is an integer), whereas its conjugated field $\theta_s$ is completely uncertain. Correlators of the charge fields remain the same as in ordinary LL. The correlators $s_{xx}(x,\tau)$ and $s_{yy}(x,\tau)$ decay exponentially to zero being averaged with the oscillating factor $e^{i\theta_s}$. But $s_{zz}(x,\tau)$ survives since $\theta_s$ doesn't appear in its expression:
\begin{align}\label{dc for SDW}
&s_{zz}(x,\tau)=\frac{2}{(\pi a_0)^2}\cos(2k_Fx)(\frac{a_0}{\sqrt {x^2+y_c^2}})^{K_c}.
\end{align}
It is determined exclusively by the charge degrees of freedom. It oscillates with the wave vector $2k_F$ and decays power-like with $\sqrt{x^2+y_c^2}$. For $K_c\approx1$ it decays as $x^{-1}$ which is slower than $x^{-2}$ decay of the ordinary LL case. This is the result of ordering in the SDW state.

\textit{Total spin correlations.---}We aim to obtain the Fourier transforms of the two-time total spin correlators. Eqs. \eqref{dc for oLL} and \eqref{dc for SDW} present the time-ordered spin density correlators for imaginary time $\tau$. The imaginary-time-ordered total spin correlators read $S^T_{aa}(\tau)=\int_0^l\int_0^ldxdx's_{aa}(x-x',\tau)\approx l \int_{-\infty}^{\infty}s_{aa}(x,\tau)$. Their Fourier transforms are $S^T_{aa}(\omega)=\int_{-\infty}^\infty e^{i\omega\tau}S^T_{aa}(\tau)$. The Fourier transform of the retarded correlator $S^R_{aa}(t)$ is related to the time-ordered one as analytic continuation: $S^R_{aa}(\omega)=S^T_{aa}(i\omega\rar \omega+i\delta)$, where $\delta=0^+$ \cite{1d book}. Details of calculation see in \cite{sm}.

However, when integrating the correlator $s_{zz}(x,\tau)$ over $x$, we are faced with the fact that, in the absence of the transverse field, the integral is not constant in time in contradiction with the exact conservation of the $z$-component of the total spin. For the SDW state $S^R_{zz}(\omega)$ is also not a constant, but the SDW appears only in non-zero transverse field that violates the $S_z$ conservation. Such a contradiction was first noted by Tennant et al. \cite{tennant} (see their appendix) and they treated it phenomenologically assuming that the oscillating term is a complete derivative.

This discrepancy originates from filling of infinite Fermi sea, a crucial assumption in the LL model \cite{Luttinger,Luttinger correction,1d book, bosonization book}. Electron and hole excitations in this model are completely symmetric. In real wires the relativistic particle-hole symmetry is violated. In particular, the momenta of holes cannot exceed $k_F$ by modulus. This limitation is not important if essential for a problem momenta are close to $\pm k_F$. This is the case for the spin-Peierls instability leading to the appearance of the SDW. However, the momenta far from $k_F$ bring a significant contribution to the total spin. Therefore, the LL model does not respect the total spin conservation. Nevertheless, calculations for the non-interacting case within the Fermi gas model shows that the cut-off of the integration at some negative moment $k_D$ leads to conserving $S_z$ if $k_D<k_F$.\cite{sm} This cut-off produces additional terms in the spin density correlator so that at $k_D=0$,
\begin{align}\label{corrected ni}
s_{zz}(x,\tau)=\frac{1}{\pi^2}\frac{y^2-x^2}{(x^2+y^2)^2}+\frac{1}{\pi^2}\frac{\cos( 2k_F x)}{x^2+y^2}-\frac{2}{\pi^2}\frac{y (y\cos(k_F x)- x\sin(k_F x))e^{ -k_F y}}{(x^2+y^2)^2},
\end{align}
where $y=v_F\tau$. The third term in Eq. \eqref{corrected ni} is the cut-off correction. After integration over $x$ it completely cancels the contribution of the second term. The first term is contribution of small momentum transfer. Its integration gives zero.

Unfortunately, it is not clear how to introduce the proper cut-off in LL model. The conjectured form of $s_{zz}(x,\tau)$ is given by Eq. \eqref{conjecture of correction} in \cite{sm}; it tends to the exact free electron correlator of Eq. \eqref{corrected ni} at vanishing interaction and, after integration over $x$, the correction approximately cancels the term with the transfer of momentum by $2k_F$. Anyway, the result \eqref{corrected ni} obtained for free electrons shows that, at $b=0$, at a proper cut-off the contribution of the $2k_F$ momentum transfer to $S_{zz}$ exactly vanishes. The same is correct in the presence of $b_\parallel$. %longitudinal field.
In the presence of non-zero $b_\perp$ its smallness is determined by the smallness of $b_\perp$. %However, its smallness is determined not only by the smallness of $b_\perp$, but also it includes a small exponent $\exp(-Cv_F/\alpha)$, where $C$ is of the order of unity.
Further we neglect this part of the correlator. For the same reason we neglect the contribution of the $2k_F$-transfer of momentum to the transverse spin correlators. Contribution from the small momentum transfer must be retained. We then arrive at a simple result for the LL state:
\begin{align}\label{new LL total}
&S^R_{xx}(\omega)=S^R_{yy}(\omega)=A_0[\omega_s^2+(\omega+i\delta)^2][\omega_s^2-(\omega+i\delta)^2]^{\frac{K_s}{2}+\frac{1}{2K_s}-2},\quad S^R_{zz}(\omega)=0.
\end{align}
where $\omega_s=2m\alpha v_s$ and $A_0=\frac{l(\frac{a_0}{2v_s})^{K_s+\frac{1}{K_s}-2}\Gamma(2-\frac{K_s}{2}-\frac{1}{2K_s})}{\pi v_s\Gamma(1+\frac{K_s}{2}+\frac{1}{2K_s})}$. The SDW state that appears only in the transverse field violating the total spin conservation does not require such a fine tuning. Its total spin correlators are:
\begin{align}\label{SDW total}
&S^R_{xx}(\omega)=S^R_{yy}(\omega)=0,\quad S^R_{zz}(\omega)=A_{SDW}[\omega_{0c}^2-(\omega+i\delta)^2]^{\frac{K_c}{2}-1},
\end{align}
where $\omega_{0c}=2k_Fv_c$, and $A_{SDW}=\frac{2l(\frac{a_0}{2v_c})^{K_c-2}\Gamma(1-\frac{K_c}{2})}{\pi v_c\Gamma(\frac{K_c}{2})}$.

\textit{Relation to experiment.---}The results given by Eqs. \eqref{new LL total} and \eqref{SDW total} show that measurements of the total spin correlators can be used as a diagnostic tool for identification of the state of the electronic liquid in the quantum wire, is it the LL or the SDW. Besides of that we predict that in the ordinary LL state the transverse correlators display the spin resonance at $\omega_s=2m\alpha v_s \approx 2k_F\alpha$. For $K_s\approx 1$ the shape of the resonance line is almost Lorentzian \cite{spin resonance}. The position of resonance agrees with the previous non-interacting result Eq. \eqref{non-int}. In the SDW state only the $z$ correlator survives and it has a peak at a relatively high $\omega_{0c}=2k_Fv_c\approx 2 k_Fv_F$.  A typical value for this frequency in semiconductors is $10^{14}$ Hz. At much lower frequency it is almost constant.

Experimentally, the Faraday rotation method \cite{Crooker 2010,Crooker} measures directly the spin correlations in real time. At zero field the system is in the ordinary LL state, and we expect peaks at $\omega=2m\alpha v_s$ for directions perpendicular to the RDI axis. The direction of the RDI axis is not a priori known. It must be found utilizing the $U(1)$ symmetry of the transverse spin correlations. Applying the magnetic field perpendicular to the RDI axis, one can check whether the wire transits to the SDW state. At this transition the longitudinal correlator suppressed in the LL state becomes dominant, whereas the transverse correlators are suppressed.

%Finally it is worth pointing out that
The considered quantum wire problem is closely related to a quantum antiferromagnetic spin chain problem, where Dzyaloshinskii-Moriya interactions plays a similar role as RDI.\cite{Starykh} Thus, studies on spin chain systems, e.g. \cite{Starykh 2011,Halg 2014} may also be helpful for understanding the physics of quantum wires. %(...)%Thus, experiments on spin chain systems, e.g. \cite{Starykh 2011,Halg 2014} may also be helpful for understanding the physics of quantum wires.

\textit{Conclusions.---}We calculated the spin density and total spin correlators in the quantum wire with RDI in the ordinary LL state and in the SDW state. They display different dependencies on directions and different positions of resonance peaks. Thus, experimental studies of spin correlations in quantum wires can be employed for detecting the SDW driven by properly directed magnetic field and electron resonance on the intrinsic field induced by the RDI. The impurity scattering does not change the results significantly if the mean free path is larger than $1/(m\alpha)$, typically 10-30 nm. The corresponding mobility is $\sim$(1-3)$\times 10^3$ cm$^2$/(Vs).

We thank Oleg A. Starykh, Nikolai A. Sinitsyn and Fuxiang Li for helpful discussions of theoretical problem and experimental situation. %We also thank Oleg A. Starykh and Nikolai A. Sinitsyn for useful comments on the manuscript....

\newpage

\section*{Supplemental Material}

\subsection{BOSONIZATION}
Here we present the procedure of bosonization in some details. We follow prescriptions given in Ref. \cite{Starykh} in the main text. The chiral fermionic fields are defined as: $R_\sigma(x)=\int\frac{dk}{2\pi}e^{i(k-k_{\sigma+})x}a_\sigma(k)$, $L_\sigma(x)=\int\frac{dk}{2\pi}e^{i(k-k_{\sigma-})x}a_\sigma(k)$, where $a_{\sigma}(k)$ is the Fermi annihilation operator in momentum space.
The second quantized wave-function operator $\Psi_{\sigma}(x)= e^{ik_{\sigma+} x}R_\sigma(x)+ e^{ik_{\sigma-} x}L_\sigma(x)$. The interaction Hamiltonian $H_{int}$ contains several quartic products of fermionic chiral fields. We neglect the strongly oscillating terms like $e^{i(k_{\sigma'-}-k_{\sigma'+}) x'}R^\dag_\sigma(x)R_\sigma(x)R^\dag_{\sigma'}(x')L_{\sigma'}(x)$. By assumption, $U(x-x')$ decreases rapidly beyond the effective interaction radius, whereas the fields $R_\sigma(x)$, $L_\sigma(x)$ vary on much longer scales. Therefore, it is possible to integrate first over the difference $x-x'$ neglecting the change of the chiral fields. After these simplifications we obtain:
\begin{align}
&H_{int}=\frac{1}{2}\sum_{\sigma,\sigma'}\int dx\{\tilde U(0)(R^\dag_\sigma R_\sigma+L^\dag_\sigma L_\sigma)(R^\dag_{\sigma'} R_{\sigma'}+L^\dag_{\sigma'}L_{\sigma'})\nn\\
&+[\tilde U(2k_F+\frac{(\sigma+\sigma')b_\parallel}{ v_F}) e^{i\frac{2(\sigma'-\sigma)b_\parallel}{ v_F} x}R^\dag_\sigma L_\sigma L^\dag_{\sigma'} R_{\sigma'}+h.c.]\},
\end{align}
where $\tilde U(q)$ is the Fourier transform of $U(x)$. We have dropped the arguments of the chiral fields while keeping in mind that coordinate of the first two fields in any term is $x$ and that of the last two is 0.

The chiral fermionic fields are now expressed in terms of chiral bosonic fields $\phi_{R/L_\sigma}$ as $R_\pm=\frac{\eta_\pm}{\sqrt{2\pi a_0}}e^{i\sqrt{4\pi}\phi_{R_\pm}}$, $ L_\pm=\frac{\eta_\pm}{\sqrt{2\pi a_0}}e^{-i\sqrt{4\pi}\phi_{L_\pm}}$, where $a_0$ is the ultraviolet cut-off,
and $\eta_\pm$ are the so-called Klein operators. The bosonic fields obey commutation relations: $[\phi_{R_\sigma},\phi_{L_{\sigma'}}]=\frac{i}{4}\delta_{\sigma\sigma'}$, $[\phi_{R/L_\sigma}(x),\phi_{R/L_{\sigma'}}(y)]=\pm\frac{i}{4}\delta_{\sigma\sigma'}\sgn(x-y)$. The Klein operators $\eta_{\pm}$ can be viewed as Majorana fermions which satisfy: $\{\eta_\sigma,\eta_{\sigma'}\}=2\delta_{\sigma\sigma'}$, $\eta_\sigma^\dag=\eta_\sigma$, $\eta_+\eta_-=i$. The commutation relations of bosonic fields ensure anticommutation relations of chiral fermionic fields with the same spin index $\sigma$. But commutators of bosonic fields between different spin species always vanish, so to ensure anticommutations between fermionic fields with different $\sigma$s the Klein operators must be introduced. The chiral densities are $R^\dag_\sigma R_\sigma=\frac{\partial_x\phi_{R_\sigma}}{\sqrt\pi}$, $L^\dag_\sigma L_\sigma=\frac{\partial_x\phi_{L_\sigma}}{\sqrt\pi}$. Finally, following general rules \cite{1d book} we introduce charge fields $\phi_{c}$, $\theta_{c}$ and spin fields $\phi_{s}$, $\theta_{s}$ related to the chiral fields as: $\phi_{R_\sigma}=\frac{(\phi_c-\sigma\phi_s)-(\theta_c-\sigma\theta_s)}{2\sqrt2}$, $ \phi_{L_\sigma}=\frac{(\phi_c-\sigma\phi_s)+(\theta_c-\sigma\theta_s)}{2\sqrt2}$. Plugging these expressions into the Hamiltonian with a careful usage of the commutations we arrive at the bosonized Hamiltonian Eq. (5) in the main text.

\subsection{CALCULATION OF SPIN DENSITY CORRELATIONS}
Here we present calculations of the spin density correlations for the ordinary LL state.
In terms of the bosonic fields, the spin density operators reads:
\begin{align}\label{spin density}
&s_x(x)=\frac{1}{\pi a_0}\sin[\sqrt{2\pi}(\theta_s-\phi_s)-2m\alpha x]+\frac{1}{\pi a_0}\sin[\sqrt{2\pi}(\theta_s+\phi_s)-2m\alpha x]\nn\\
&+\frac{1}{\pi a_0}\sin[\sqrt{2\pi}(\theta_s-\phi_c)+(2k_F-2m\alpha)x]+\frac{1}{\pi a_0}\sin[\sqrt{2\pi}(\theta_s+\phi_c)-(2k_F+2m\alpha)x],\nn\\
&s_y(x)=\frac{1}{\pi a_0}\cos[\sqrt{2\pi}(\theta_s-\phi_s)-2m\alpha x]+\frac{1}{\pi a_0}\cos[\sqrt{2\pi}(\theta_s+\phi_s)-2m\alpha x]\nn\\
&+\frac{1}{\pi a_0}\cos[\sqrt{2\pi}(\theta_s-\phi_c)+(2k_F-2m\alpha)x]+\frac{1}{\pi a_0}\cos[\sqrt{2\pi}(\theta_s+\phi_c)-(2k_F+2m\alpha)x],\nn\\
&s_z(x)=-\sqrt{\frac{2}{\pi}}\partial_x\phi_s(x)-\frac{1}{\pi a_0}\sin[\sqrt{2\pi}(\phi_c-\phi_s)+2k_Fx]+\frac{1}{\pi a_0}\sin[\sqrt{2\pi}(\phi_c+\phi_s)+2k_Fx].
\end{align}
Let us define the partition function as a functional integral:
\begin{equation}\label{}
Z=\int \mathcal{D}\Phi(x,\tau) e^{\int_0^{\beta}d\tau\int dx\mathcal{L}(\Phi(x,\tau))},
\end{equation}
where $\tau=it+\epsilon \sgn(t)$($\epsilon=0^+$) is the imaginary time, $\beta=1/(k_BT)$, $\Phi=(\phi_c,\theta_c,\phi_s,\theta_s)$ is the 4-vector of fields, and $\mathcal{L}(\Phi(x,\tau))$ is the Lagrangian associated with the Hamiltonian $H$. Note that for the ordinary LL state $H$ is completely quadratic and thus invariant under a uniform translation of any bosonic fields: $\Phi_i(x)\rar\Phi_i(x)+A_i$, a symmetry which we use later. In the functional integral language, the time-ordered correlation for operators $ A(\Phi)$ and $ B(\Phi)$ is:
\begin{equation}\label{}
\la\operatorname{T_\tau}A(\tau)B(0)\ra=\frac{1}{Z}\int \mathcal{D}\Phi(x,\tau) A(\Phi(\tau))B(\Phi(0)) e^{\int_0^{\beta}d\tau\int dx\mathcal{L}(\Phi(x,\tau))}.
\end{equation}
Later we will drop the time ordering symbol $\operatorname {T_{\tau}}$ and use directly $\la\ra$ to denote the time-ordered average. The Lagrangian $\mathcal{L}$ can be written as $\mathcal{L}(\Phi)=-\frac{1}{2}\Phi M\Phi=\frac{1}{2}\Phi_i M_{ij}\Phi_j$, where the Fourier transform of the matrix $M(x,\tau)$ is:
\begin{equation}
M(q,\omega)=\left(\begin{array}{cccc}
\frac{v_cq^2}{K_c} & iq\omega & 0 &  0\\
iq\omega  & v_cK_cq^2 & 0 & 0\\
0 & 0 & \frac{v_sq^2}{ K_s} &  iq\omega\\
0 & 0 & iq\omega  & v_sK_sq^2
\end{array}\right).
\end{equation}
Note that here $\omega$ is the imaginary frequency associated with $\tau$. The inverse of $M(q,\omega)$ reads:
\begin{equation}\label{}
M^{-1}(q,\omega)=\left(\begin{array}{cccc}
\frac{ K_cv_c}{\Omega_c^2} & -\frac{i\omega}{q\Omega_c^2} & 0 & 0\\
-\frac{i\omega}{q\Omega_c^2}  & \frac{ v_c}{ K_c\Omega_c^2} & 0 & 0\\
0 & 0 & \frac{ K_sv_s}{\Omega_s^2} & -\frac{i\omega}{q\Omega_s^2}\\
0 & 0 &  -\frac{i\omega}{q\Omega_s^2} & \frac{ v_s}{ K_s\Omega_s^2}
\end{array}\right),
\end{equation}
where we denoted $\Omega_{c/s}^2=v_{c/s}^2q^2+\omega^2$. Let $\Phi_i(q,\omega)$ be the Fourier transform of $\Phi_i(x,\tau)$. Then:
\begin{equation}\label{formula}
\la\Phi_i(q,\omega)\Phi_j(-q,-\omega)\ra=\beta lM^{-1}_{ij}(q,\omega).
\end{equation}
Correlations for $\phi_{s/c}(x,\tau)$ and $\theta_{s/c}(x,\tau)$ can be obtained from Eq. \eqref{formula} by inverse Fourier transform. The results at zero temperature are:
\begin{subequations}
\begin{align}
&\la(\phi_{s/c}(x,\tau)-\phi_{s/c}(0,0))^2\ra=\frac{K_{s/c}}{2\pi}\log\frac{x^2+y_{s/c}(\tau)^2}{a_0^2},\label{formula1} \\
&\la(\theta_{s/c}(x,\tau)-\theta_{s/c}(0,0))^2\ra = \frac{1}{2\pi K_{s/c}}\log\frac{x^2+y_{s/c}(\tau)^2}{a_0^2},\label{formula2}\\
&\la\phi_{s/c}(x,\tau)\theta_{s/c}(0,0)\ra= -\frac{i}{2\pi}\mathrm{Arg}[y_{s/c}(\tau)+ix],\label{formula3}
\end{align}
\end{subequations}
where $y_{s/c}(\tau)=v_{s/c}\tau+a_0\sgn(\tau)$. The argument in Eq. \eqref{formula3} is defined with a branch cut at $(-\infty,0]$.

When calculating the spin density correlators $s_{aa}(x,t)$ employing Eq. \eqref{spin density}, there appear terms of three types:
(a) $\la\partial_x\phi_s(x,\tau)\partial_{x}\phi_s(0,0)\ra$, (b) $\la\partial_x\phi_s(x,\tau)e^{i\sum A_i\Phi_i(0,0)}\ra$, and (c) $\la e^{i\sum B_i\Phi_i(x,\tau)}e^{i\sum C_i\Phi_i(0,0)}\ra$, where $A_i, B_i, C_i$ are numerical coefficients. For their calculation we employ the invariance of $H$ and $\mathcal{L}$ under the uniform translation of $\Phi_i$. For terms of type (b) with $A_i\neq 0$, the translation $\Phi_i\rightarrow \Phi_i+\pi/A_i$ changes the sign of the averaged value leaving the Lagrangian invariant. Thus, the average $\la\partial_x\phi_s(x,\tau)e^{i\sum A_i\Phi_i(0,0)}\ra$ must be zero if at least one $A_i\neq 0$. For terms of type (c), a similar argument shows that $\la e^{i\sum B_i\Phi_i(x,\tau)}e^{i\sum C_i\Phi_i(0,0)}\ra=0$ if at least one of the sums $B_i+C_i\neq 0$. As a result, $s_{zz}(x,\tau)$ reduces to:
\begin{align}\label{}
&s_{zz}(x,\tau)=\frac{2}{\pi}\la\partial_x\phi_s(x,\tau)\partial_{x}\phi_s(0,0)\ra\nn\\
&+\frac{1}{4\pi^2a_0^2}[e^{i2k_Fx}\la e^{i\sqrt{2\pi}(\phi_c(x,\tau)-\phi_s(x,\tau))}e^{-i\sqrt{2\pi}(\phi_c(0,0)-\phi_s(0,0))}\ra+h.c.]\nn\\
&+\frac{1}{4\pi^2a_0^2}[e^{i2k_Fx}\la e^{i\sqrt{2\pi}(\phi_c(x,\tau)+\phi_s(x,\tau))}e^{-i\sqrt{2\pi}(\phi_c(0,0)+\phi_s(0,0))}\ra+h.c.].
\end{align}
From Eq. \eqref{formula1} it follows that
\begin{align}\label{35}
&\frac{2}{\pi}\la\partial_x\phi_s(x,\tau)\partial_{x}\phi_s(0,0)\ra=\frac{2}{\pi}\partial_x\partial_{x'}\la-\frac{1}{2}(\phi_s(x,\tau)-\phi_s(x',0))^2\ra|_{x'=0}\nn\\
&=\frac{2}{\pi}\partial_x\partial_{x'}[-\frac{K_s}{4\pi}\log[\frac{(x-x')^2+y_s(\tau)^2}{a_0^2}]]|_{x'=0}=\frac{K_s}{\pi^2}\frac{y_s^2-x^2}{(x^2+y_s^2)^2}.
\end{align}
For the second term in Eq. \eqref{35} we apply the formula $\la e^{i A}\ra=e^{-\frac{1}{2}\la A^2\ra}$ valid for any Gaussian distributed variable $A$. Let calculate for example an average:
\begin{align}\label{}
&\la e^{i\sqrt{2\pi}(\phi_c(x,\tau)-\phi_s(x,\tau))}e^{-i\sqrt{2\pi}(\phi_c(0,0)-\phi_s(0,0))}\ra\nn\\
&=\la e^{i\sqrt{2\pi}(\phi_c(x,\tau)-\phi_s(x,\tau))-i\sqrt{2\pi}(\phi_c(0,0)-\phi_s(0,0))}\ra=e^{-\pi\la[\phi_c(x,\tau)-\phi_s(x,\tau)-\phi_c(0,0)+\phi_s(0,0)]^2\ra}\nn\\
&=e^{-\pi[\frac{K_{c}}{2\pi}\log\frac{x^2+y_{c}(\tau)^2}{a_0^2}+\frac{K_{s}}{2\pi}\log\frac{x^2+y_{s}(\tau)^2}{a_0^2}]}=(\frac{a_0}{\sqrt{x^2+y_c^2}})^{K_c}(\frac{a_0}{\sqrt{x^2+y_s^2}})^{K_s}.
\end{align}
Similar calculations can be done for other terms in $s_{zz}(x,\tau)$ and for the other two spin density correlators, which lead to the results Eq. \eqref{dc for oLL} in the main text. The $z$ direction spin density correlators of the SDW state can also be calculated in the same way, with $\phi_s$ replaced by a constant that minimizes $H_C$.

\subsection{INTEGRALS OF TOTAL SPIN CORRELATIONS}
We calculate from the spin density correlators the total spin correlators by integrate over the coordinate. The integrals that must be evaluated are of the forms:
\begin{align}
&I_1=\int_{-\infty}^\infty d\tau  \int_{-\infty}^\infty dxe^{i\omega\tau}\cos(kx)\frac{y_s^2-x^2}{(x^2+y_s^2)^a},\nn\\
&I_2=\int_{-\infty}^\infty d\tau  \int_{-\infty}^\infty dx e^{i\omega\tau}\cos(kx)\frac{1}{(x^2+y_c^2)^b(x^2+y_s^2)^c},\nn\\
&I_3=\int_{-\infty}^\infty d\tau  \int_{-\infty}^\infty dx e^{i\omega\tau}\cos(kx)\frac{1}{(x^2+y_c^2)^d},
\end{align}
where $k\geq 0$ and $a,b,c$ are constants, and $\omega$ is imaginary frequency associated with $\tau$. Of them, $I_1$ or $I_2$ are parts of the correlations with small $q$ or $q\sim 2k_F$ of the ordinary LL state, and $I_3$ corresponds to the ($z$ component) correlation of the SDW state.

The integrals can be performed by changing to polar coordinates $(r,\phi)$ where $x = r \cos \phi$ and $y_s = r \sin \phi$. For example, $I_1$ reads:
\begin{align}
&I_1=\frac{1}{v_s}\int_0^{2\pi}d\phi\int_{0}^\infty dr e^{ir(k \cos\phi+\frac{\omega}{v_s}\sin\phi)}\frac{r^2(\sin^2\phi-\cos^2\phi)}{r^{2a-1}}\nn\\
&=-\frac{1}{v_s}\int_0^{2\pi}d\phi\int_{0}^\infty dr e^{ir\sqrt{k^2+\frac{\omega^2}{v_s^2}}\cos(\phi-\arctan\frac{\omega}{kv_s})}\frac{\cos(2\phi)}{r^{2a-3}}\nn\\
&=\frac{\pi\Gamma(3-a)}{ v_s\Gamma(a)}\frac{k^2v_s^2-\omega^2}{k^2v_s^2+\omega^2}(\frac{k^2v_s^2+\omega^2}{4v_s^2})^{a-2}.
\end{align}

Similarly we can calculate $I_3$, and also $I_2$ provided we approximate both $v_c$ and $v_s$ to be $v_F$. The results are:
\begin{align}
&I_2=\frac{\pi \Gamma(1-b-c)}{v_F\Gamma(b+c)}(\frac{k^2v_F^2-\omega^2}{4v_F^2})^{b+c-1},\nn\\
&I_3=\frac{\pi \Gamma(1-d)}{v_c\Gamma(d)}(\frac{k^2v_c^2-\omega^2}{4v_c^2})^{d-1}.
\end{align}

Applying these results to the correlations and analytically continuing to real frequency by $i\omega\rar\omega+i\delta$, we obtain the correlations for the ordinary LL state:
\begin{align}
&S^R_{xx}(\omega)=S^R_{yy}(\omega)\nn\\
&=A_0(\omega_s^2+\omega^2)(\omega_s^2-\omega^2)^{\frac{K_s}{2}+\frac{1}{2K_s}-2}+\frac{1}{2}A^x_{2k_F}[(\omega_{+}^2-\omega^2)^{\frac{K_c}{2}+\frac{1}{2K_s}-1}+(\omega_-^2-\omega^2)^{\frac{K_c}{2}+\frac{1}{2K_s}-1}],\nn\\
&S^R_{zz}(\omega)=A^z_{2k_F}(\omega_0^2-\omega^2)^{\frac{K_c}{2}+\frac{1}{2K_s}-1},
\end{align}
and those for the SDW state:
\begin{align}\label{}
&S^R_{xx}(\omega)=S^R_{yy}(\omega)=0,\quad S^R_{zz}(\omega)=A_{SDW}(\omega_{0c}^2-\omega^2)^{\frac{K_c}{2}-1},
\end{align}
where we defined the frequencies $\omega_s=2m\alpha v_s$, $\omega_0=2k_Fv_F$, $\omega_\pm=2(k_F\pm m\alpha ) v_F$, $\omega_{0c}=2k_Fv_c$, and the amplitudes
\begin{align}
&A_0=\frac{l(\frac{a_0}{2v_s})^{K_s+\frac{1}{K_s}-2}\Gamma(2-\frac{K_s}{2}-\frac{1}{2K_s})}{\pi v_s\Gamma(1+\frac{K_s}{2}+\frac{1}{2K_s})},\nn\\
&A^x_{2k_F}=\frac{l(\frac{a_0}{2v_F})^{K_c+\frac{1}{K_s}-2}\Gamma(1-\frac{K_c}{2}-\frac{1}{2K_s})}{\pi v_F\Gamma(\frac{K_c}{2}+\frac{1}{2K_s})},\nn\\
&A^z_{2k_F}=\frac{l(\frac{a_0}{2v_F})^{K_c+K_s-2}\Gamma(1-\frac{K_c}{2}-\frac{K_s}{2})}{\pi v_F\Gamma(\frac{K_c}{2}+\frac{K_s}{2})},\nn\\
&A_{SDW}=\frac{2l(\frac{a_0}{2v_c})^{K_c-2}\Gamma(1-\frac{K_c}{2})}{\pi v_c\Gamma(\frac{K_c}{2})}.
\end{align}
$\omega$ everywhere in these expressions is understood to have a small imaginary part. Note that the expressions for the $q\sim 2k_F$ parts of the ordinary LL correlations are only approximations when $v_c$ and $v_s$ are both close to $v_F$.

We remind that the conservation of $S_z$ requires that $S_{zz}^R(\omega)=0$ at any $\omega$ in the absence of the transverse external magnetic field.  In the main text we have demonstrated that this discrepancy is associated with the inconsistency of the LL model at negative, large by modulus $k$ and how this discrepancy can be corrected.
\subsection{NON-INTERACTING MODEL WITH VARYING FILLING DEPTH OF FERMI SEA}
We calculate the spin density correlators for the non-interacting case with no RDI and zero external field to illustrate the effect of filling of Fermi sea on the spin correlations. The original quadratic spectrum is $E_k=k^2/(2m)$, where in the ground state the momentum states $k\in (-k_F,k_F)$ are occupied. These states all have non-negative energies. But in the LL model, the spectrum is linearized in such a way that the ground state is a filled Fermi sea with infinite depth. The spectrum is $E_k=\pm v_F k$ for right and left movers, respectively.

We will consider a more general model with a cut-off $k_D$: in the ground state the occupied states are $k\in (-k_D,k_F)$ for right movers, and $k\in (-k_F,k_D)$ for left movers (See Fig. 1). The parameter $k_D$ denotes the depth of the Fermi sea: the lowest occupied level of each species of movers has energy $- v_F k_D$. In this model, the total number of electrons is finite at any finite $k_D$. $k_D=0$ corresponds to the case when there's only one band with non-negative energy states, which is the case of the original model where electrons fill from zero energy to the Fermi surface. $k_D=\infty$ corresponds to the case of filling an infinite sea and of infinite number of particles, as assumed in the LL model.
\begin{figure}
  \centering
  \includegraphics[width=0.5\textwidth]{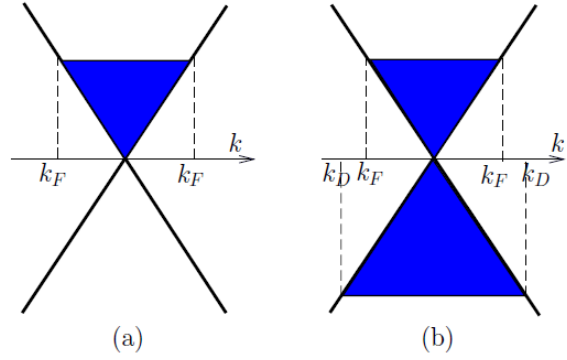}\\
  \caption{Spectra: (a) for the original model; (b) for the generalized model.}\label{19}
\end{figure}
The field operator $\Psi_\sigma(x,t)$ reads:
\begin{equation}\label{1}
\Psi_\sigma(x,t)=e^{i k_F x}R_\sigma(x,t)+e^{-i k_F x}L_\sigma(x,t),
\end{equation}
where $R_\sigma(x,t)$ and $L_\sigma(x,t)$ are the chiral fields. The Fourier expansion of the chiral field contain the operators of annihilation in momentum space as Fourier coefficients:
\begin{align}\label{2}
&R_\sigma(x,t)=\frac{1}{\sqrt{l}}\sum_{k}a_{k,\sigma}e^{i(kx-kv_Ft)},\nn\\
&L_\sigma(x,t)=\frac{1}{\sqrt{l}}\sum_{k}b_{k,\sigma}e^{i(kx+kv_Ft)}.
\end{align}
Since the factors $e^{\pm ik_F x}$ are placed in front of sums in Eq. \eqref{2}, the summation proceeds from $-k_F-k_D$ till zero for right movers and from 0 till $k_F+k_D$ for left movers. At zero temperature the momentum space field operators satisfy:
\begin{align}\label{pairing}
&\la a^\dag_{k_1,\sigma_1}a_{k_2,\sigma_2}\ra=\delta_{k_1,k_2}\delta_{\sigma_1,\sigma_2}\Theta(-k_1)Y(k_1+k_F+k_D),\nn\\
&\la a_{k_1,\sigma_1}a^\dag_{k_2,\sigma_2}\ra=\delta_{k_1,k_2}\delta_{\sigma_1,\sigma_2}\Theta(k_1),\nn\\
&\la b^\dag_{k_1,\sigma_1}b_{k_2,\sigma_2}\ra=\delta_{k_1,k_2}\delta_{\sigma_1,\sigma_2}\Theta(k_1)Y(-k_1+k_F+k_D),\nn\\
&\la b_{k_1,\sigma_1}b^\dag_{k_2,\sigma_2}\ra=\delta_{k_1,k_2}\delta_{\sigma_1,\sigma_2}\Theta(-k_1).
\end{align}
The averages of all other pairings between $a$, $a^\dag$, $b$, $b^\dag$ vanish. The spin density operators are $s_a(x,t)=\Psi^\dag_\sigma(x,t)\sigma_{a,\sigma\sigma'}\Psi_{\sigma'}(x,t)$. The correlator of $z$-components of spin at positive time reads:
\begin{align}\label{}
&s_{zz}(x,t)=\la s_z(x,t)s_z(0,0)\ra\nn\\
&=\la\Psi^\dag_{+}(x,t)\Psi_{+}(x,t)\Psi^\dag_{+}(0,0)\Psi_{+}(0,0)\ra
+\la\Psi^\dag_{-}(x,t)\Psi_{-}(x,t)\Psi^\dag_{-}(0,0)\Psi_{-}(0,0)\ra\nn\\
&-\la\Psi^\dag_{+}(x,t)\Psi_{+}(x,t)\Psi^\dag_{-}(0,0)\Psi_{-}(0,0)\ra
-\la\Psi^\dag_{-}(x,t)\Psi_{-}(x,t)\Psi^\dag_{+}(0,0)\Psi_{+}(0,0)\ra.
\end{align}
We then apply Wick's theorem and express each term as a sum of all possible pairings of operators. Due to the reflection symmetry of the spin space, $\la\Psi^\dag_{+}(x,t)\Psi_{+}(x,t)\ra=\la\Psi^\dag_{-}(x,t)\Psi_{-}(x,t)\ra$. Thus,
\begin{align}\label{}
&s_{zz}(x,t)=2\la\Psi^\dag_{+}(x,t)\Psi_{+}(0,0)\ra\la\Psi_{+}(x,t)\Psi^\dag_{+}(0,0)\ra.
\end{align}
In terms of chiral fields we find:
\begin{align}\label{7}
&s_{zz}(x,t)=2(e^{-i k_F x}\la R^\dag_+(x,t)R_+(0,0)\ra+e^{i k_F x}\la L^\dag_+(x,t)L_+(0,0)\ra)\nn\\
&\times(e^{i k_F x}\la R_+(x,t)R^\dag_+(0,0)\ra+e^{-i k_F x}\la L^\dag_+(x,t)L^\dag_+(0,0)\ra).
\end{align}
The chiral field averages are readily found employing Eqs. \eqref{2} and \eqref{pairing}:
\begin{align}\label{}
&\la R^\dag_+(x,\tau)R_+(0,0)\ra=\frac{1}{2\pi}\frac{1-e^{-(k_F+k_D)(-ix+v_F\tau)}}{-ix+v_F\tau},\nn\\
&\la L^\dag_+(x,\tau)L_+(0,0)\ra=\frac{1}{2\pi}\frac{1-e^{-(k_F+k_D)(ix+v_F\tau)}}{ix+v_F\tau},\nn\\
&\la R_+(x,\tau)R^\dag_+(0,0)\ra=\frac{1}{2\pi}\frac{1}{-ix+v_F\tau},\nn\\
&\la L_+(x,\tau)L^\dag_+(0,0)\ra=\frac{1}{2\pi}\frac{1}{ix+v_F\tau},
\end{align}
where we assumed $\tau>0$. Note that if $k_D\rar\infty$ (the LL prescription), $\la R^\dag_+(x,t)R_+(0,0)\ra=\la R_+(x,t)R^\dag_+(0,0)\ra$ and $\la L^\dag_+(x,t)L_+(0,0)\ra=\la L_+(x,t)L^\dag_+(0,0)\ra$, as a consequence of electron-hole symmetry. For finite $k_D$, electrons and holes are not symmetric. Thus, $\la R^\dag_+(x,t)R_+(0,0)\ra\neq\la R_+(x,t)R^\dag_+(0,0)\ra$.

Plugging these results into Eq. \eqref{7}, we get:
\begin{align}\label{general model result}
&s_{zz}(x,\tau)
=\frac{1}{\pi^2}\frac{v_F^2\tau^2-x^2}{(x^2+v_F^2\tau^2)^2}+\frac{1}{\pi^2}\frac{\cos( 2k_F x)}{x^2+v_F^2\tau^2}\nn\\
&-\frac{2}{\pi^2}\frac{e^{ -(k_D+k_F )v_F \tau}(v_F\tau \cos(k_D x)- x\sin(k_D x))(v_F \tau\cos(k_F x)- x\sin(k_F x))}{(v_F^2\tau^2+x^2)^2}.
\end{align}
The first and the second term in Eq. \eqref{general model result} corresponds to the contribution of small $q$ and the $q\sim 2k_F$ part, respectively (compare Eq. \eqref{dc for oLL} in the main text). The third term depends on $k_D$, and it vanishes as $k_D\rar \infty$.

The total spin correlation is obtained after integration over $x$:
\begin{align}\label{38}
&S_{zz}(\tau)=l\int_{-\infty}^{\infty}dx s_{zz}(x,\tau)=\frac{le^{ -2k_Fv_F \tau }}{\pi v_F\tau}-\frac{le^{ -2\max (k_F,k_D) v_F \tau}}{\pi v_F\tau}\nn\\
&=\left\{\begin{array}{ll}
0 & \textrm{for } 0\le k_D\le k_F,\\
\frac{l}{\pi v_F\tau}(e^{ -2k_Fv_F \tau }-e^{ -2k_Dv_F \tau }) & \textrm{for } k_D>k_F.
\end{array}\right.
\end{align}
In the second line in the equation \eqref{38}, the first term comes from the $q\sim 2k_F$ part, and the second term comes from the $k_D$-dependent part. If $0\le  k_D<k_F$ these two terms cancel each other, but if $k_D>k_F$ they do not cancel. Thus, the total $z$ spin correlation is zero if $0\le  k_D<k_F$, and it becomes non-zero when $k_D$ exceeds $k_F$. At $k_D\rar \infty$ as is in the LL model, $S_{zz}(\tau)=\frac{l}{\pi v_F\tau}e^{ -2k_Fv_F \tau }$. Therefore, it is clear that non-zero $S_{zz}$ results from inclusion of negative energy states.

Next we should extrapolate Eq. \eqref{38} to the interacting case. Corrected $s_{zz}(x,\tau)$ should satisfy: 1. its integration over x should give zero; 2. in the non-interacting limit it should reduce to Eq. \eqref{corrected ni} in the main text. A candidate could be:
\begin{align}\label{conjecture of correction}
&s_{zz}(x,\tau)=\frac{K_s}{\pi^2}\frac{(y^2-x^2)(1-e^{ -k_Fy}\cos(k_Fx))+2xye^{ -k_Fy}\sin(k_Fx)}{(x^2+y^2)^2}\nn\\
&+\frac{ a_0^{K_c+K_s-2}}{\pi^2}\frac{\cos( 2k_F x)-2^{\frac{K_c}{2}+\frac{K_s}{2}-1}e^{-k_Fy}\cos(k_Fx)}{(x^2+y^2)^{\frac{K_c}{2}+\frac{K_s}{2}}},
\end{align}
where we have approximated both $v_c$ and $v_s$ to be $v_F$. It satisfies condition 2. For condition 1, the first term integrate exactly to zero, but the correction to the second $q\sim 2k_F$ term only compensates the original result to the leading order in $1/(k_F y)$. Finally, spin density correlations of the transverse directions must be corrected in a similar manner.

\end{document}